# Pattern Matching Analysis of Electron Backscatter Diffraction Patterns for Pattern Centre, Crystal Orientation and Absolute Elastic Strain Determination – accuracy and precision assessment

Tomohito Tanaka[a,b,*], Angus J. Wilkinson[a]

[a] Department of Materials, University of Oxford, Parks Road, Oxford, OX1 3PH, UK
[b] Nippon Steel & Sumitomo Metal Corporation, 20-1 Shintomi, Futtsu, Chiba, Japan

# Pattern Matching Analysis of Electron Backscatter Diffraction Patterns for Pattern Centre, Crystal Orientation and Absolute Elastic Strain Determination – accuracy and precision assessment

Tomohito Tanaka[a,b,*], Angus J. Wilkinson[a]

[a] Department of Materials, University of Oxford, Parks Road, Oxford, OX1 3PH, UK
[b] Nippon Steel & Sumitomo Metal Corporation, 20-1 Shintomi, Futtsu, Chiba, Japan

* corresponding author.
E-mail: tomohito.tanaka@materials.ox.ac.uk

Angus Wilkinson – angus.wilkinson@materials.ox.ac.uk

**Abstract**

Pattern matching between target electron backscatter patterns (EBSPs) and dynamically simulated EBSPs was used to determine the pattern centre (PC) and crystal orientation, using a global optimisation algorithm. Systematic analysis of error and precision with this approach was carried out using dynamically simulated target EBSPs with known PC positions and orientations. Results showed that the error in determining the PC and orientation was $< 10^{-5}$ of pattern width and $< 0.01°$ respectively for the undistorted full resolution images (956×956 pixels). The introduction of noise, optical distortion and image binning was shown to have some influence on the error although better angular resolution was achieved with the pattern matching than using conventional Hough transform-based analysis. The accuracy of PC determination for the experimental case was explored using the High Resolution (HR-) EBSD method but using dynamically simulated EBSP as the reference pattern. This was demonstrated through a sample rotation experiment and strain analysis around an indent in interstitial free steel.

## Highlights:

- Systematic analysis on the efficacy of the use of dynamically simulated EBSPs for calibration and absolute strain determination was carried out

- A very accurate route for detector geometry calibration was demonstrated

- Absolute strain analysis with accuracy of the order of $10^{-4}$ is possible using dynamically simulated reference EBSPs

# 1. Introduction

High angular resolution electron backscatter diffraction (HR-EBSD) measures the small shifts caused by elastic strain and lattice rotation in the sub-regions of EBSPs between a reference pattern and test patterns using cross correlation scheme [1-3]. HR-EBSD enables the measurement of strain with a strain sensitivity of $\sim 10^{-4}$ or better at microstructure scale [1] and has been applied successfully to strain analysis in a wide range of crystalline materials including functional ceramics *e.g.* SiGe [1-4], InAlN [5], metals *e.g.* Cu [6], Fe [7-9], Ti [10], and geological materials [11, 12]. However, the quantitative analysis of elastic strain with HR-EBSD cannot be performed unless the strain state at the reference point is already known. Therefore, the application of HR-EBSD to the general case where one cannot suppose the strain state at a reference point has a substantial limitation.

To tackle this issue, the utilisation of computer simulated EBSPs as true reference patterns has been recently explored by some research groups [4, 13-16]. It is noted that when simulated EBSPs are to be used as reference patterns, the effect of uncertainty in determining PC position on strain analysis becomes non-trivial, as indicated by several authors [4, 17-20]. The uncertainty of 0.5 % of pattern width in PC determination by conventional calibration techniques [21-24] is known to introduce phantom strain of the order of $10^{-3}$ [17]. Furthermore, the presence of optical distortion and image sensor (CCD or CMOS) distortion [25] needs to be taken into account for the very accurate strain characterisation if the extent of the distortion is large enough [17].

For the accurate PC localisation for absolute strain analysis via simulation based HR-EBSD, Fullwood *et al* [14] and Alkorta *et al* [15] determined the PC position in a very similar way. First, an initial guess of the PC position and crystal orientation is given and then the corresponding simulated pattern is generated. The PC position is corrected iteratively such that the strain/crystal distortion determined via HR-EBSD between simulated and experimental EBSP is minimised or such that the extent of PC correction becomes smaller than a user set threshold. The crystal orientation can be corrected as well from the initially given orientation using the deformation gradient tensor components calculated via HR-EBSD. To produce the required large number of simulated EBSPs in a practical timescale, kinematically simulated images are used because of the lower computational demand compared to dynamical simulation. Although it was pointed out that the use of a kinematically simulated EBSP as a reference pattern can generate significant errors in the measured shifts when comparing to experimental patterns [17], Alkorta *et al* used EBSP gradient images for image correlation [15] resulting in a very precise determination of PC position and strain. Although this novel approach can avoid the significant shift measurement errors, unfortunately the error in PC determination (the difference between true PC position and measured one[1]) and the accuracy in strain measurement for all strain tensor components is not mentioned (only 2 out of 6 independent strain tensor components are shown). Furthermore, this calibration procedure requires prior knowledge of mean biaxial stress state in the analysis region through other techniques such as X-ray diffraction. On the other hand, Fullwood *et al* measured the tetragonality of SiGe semiconductor as strain through the use of kinematically simulated EBSPs [14]. All experimental EBSPs in a line scan were compared to simulated EBSPs incorporating deformation gradient tensor into the kinematical simulation. The reported difference between the expected tetragonality and measured one was $\sim 2.0 \times 10^{-3}$ with precision of $\sim 1.6 \times 10^{-3}$. Dynamically simulated EBSPs are also used as reference patterns with improving the precision, although the kinematical simulation is better in accuracy when the tetragonality is ~2 %. [16]. There is great potential of simulation based HR-EBSD for absolute strain analysis, however, more development is needed to deliver a robust method to achieve strain measurement accuracy of the order of $10^{-4}$, which is required for most target applications in metallic materials.

As to the calibration, the PC position was determined before tetragonality analysis via strain minimisation assuming that the stress along surface normal in the analysis region is null (not only $\sigma_{33}$ but also $\sigma_{13}$ and $\sigma_{23,}$ are set to be zero), and using the fixed PC position, orientation was determined separately. This is because, as Alkorta made very clear [18], it is impossible to determine PC position and displacement gradient tensor at the same time by analysing the shift in the centrelines of Kikuchi bands.

To this end, an alternative approach to determine PC position and orientation was explored in this study without relying on the local shift analysis using HR-EBSD. This would lead to further improvement in the accuracy of calibration. For this purpose, we explored the application of a global optimisation algorithm to the simultaneous determination of PC and crystal orientation by finding the best fit simulated EBSP to target EBSPs. Similar approach for calibration or solving pseudosymmetry problems can be seen in some publications [27-29], although the systematic analysis on the calibration error, accuracy and precision has not yet been reported. Global optimisation using genetic or evolution algorithms [30] can find optimal solutions with less risk of stagnating at a local minima in search space compared to gradient based optimisation schemes and it is also applicable to non-derivative function. We applied a differential evolution (DE) algorithm [30, 31] to optimise PC and orientation. The efficacy of DE to EBSD calibration and indexing was presented through the error analysis of retrieved PC and orientation for dynamically simulated target EBSPs with known PC positions and orientations. Effect of the application of noise, barrel distortion and binning to target EBSPs on the calibration and orientation error was also studied. Furthermore, the efficacy of the use of a dynamically simulated reference EBSP simulated using optimised PC and orientation to absolute elastic strain analysis was demonstrated.

---

[1] The terms, error, accuracy and precision, defined in [26] are used throughout this paper. The *error* is the difference between a true quantity and the measured one. The *accuracy* is the maximum deviation of a measured quantity from its inferred possible true value. The *precision* is the deviation between all the estimated quantities.

## 2. Methodology

### 2-1. PC and orientation optimisation procedure

For the optimisation of PC and crystal orientation (*i. e.* Euler angles), dynamically simulated EBSPs [32] generated by changing PC position and Euler angles were compared to a target EBSP in order to obtain the best matched simulated EBSP. The similarity between target and dynamically simulated EBSPs was assessed using the normalised cross correlation coefficient, *r*, which is given as:

$$r = \frac{\sum_m \sum_n (A_{mn} - \bar{A})(B_{mn} - \bar{B})}{\sqrt{(\sum_m \sum_n (A_{mn} - \bar{A})^2)(\sum_m \sum_n (B_{mn} - \bar{B})^2)}} \ , \tag{1}$$

where $m, n \in \mathbb{N}$, $A_{mn}$ and $B_{mn}$ are pixel intensities at a position (*m, n*) on EBSP images A and B. $\bar{A}$ and $\bar{B}$ correspond to the average of image intensities. One of the images is a target EBSP and the other is the dynamically simulated EBSP.

For the rapid creation of multiple dynamically simulated EBSPs, the approach used in [33] was employed such that gnomonic projected EBSPs are extracted from a master pattern. For instance, Callahan and Graef used a modified Lambert projection image as a master pattern [33]. Although the dynamical simulation of a master pattern is time consuming, once the master pattern is created the extraction of gnomonic EBSPs from it is very quick as the extraction process is just an interpolation of pixel intensities. We chose a stereographic EBSP as a master pattern. For the computer used in this study (Intel Corei7-6700K CPU 4.0 GHz), the extraction of a gnomonic EBSP with ~1000 × 1000 pixel resolution using in-house MATLAB code required only ~0.1 s. The extracted region from a master pattern is selected by designating PC position and crystal orientation, and these parameters were optimised by searching the maximum cross-correlation coefficient.

There are six parameters to be optimised (Euler angles, ($\varphi_1$, $\Phi$, $\varphi_2$), and PC $x$, $y$ and $z$ position). In this study, the reference coordinate frame is given in Fig. 1 and the Euler angles are defined as a set of three angles used to match the reference frame with the crystal frame by the passive rotation (*Z-X-Z*, Bunge notation). The PC is the foot of the normal to the phosphor screen from the source point in which diffracted electrons are generated. The position is described as ($PC_x$, $PC_y$, $PC_z$), where $PC_x$ and $PC_y$ are measured from the top left corner of the image. $PC_x$ is described as a fraction of the pattern width while $PC_y$ is a fraction of the pattern height. $PC_z$ corresponds to the camera length normalised with respect to the pattern height.

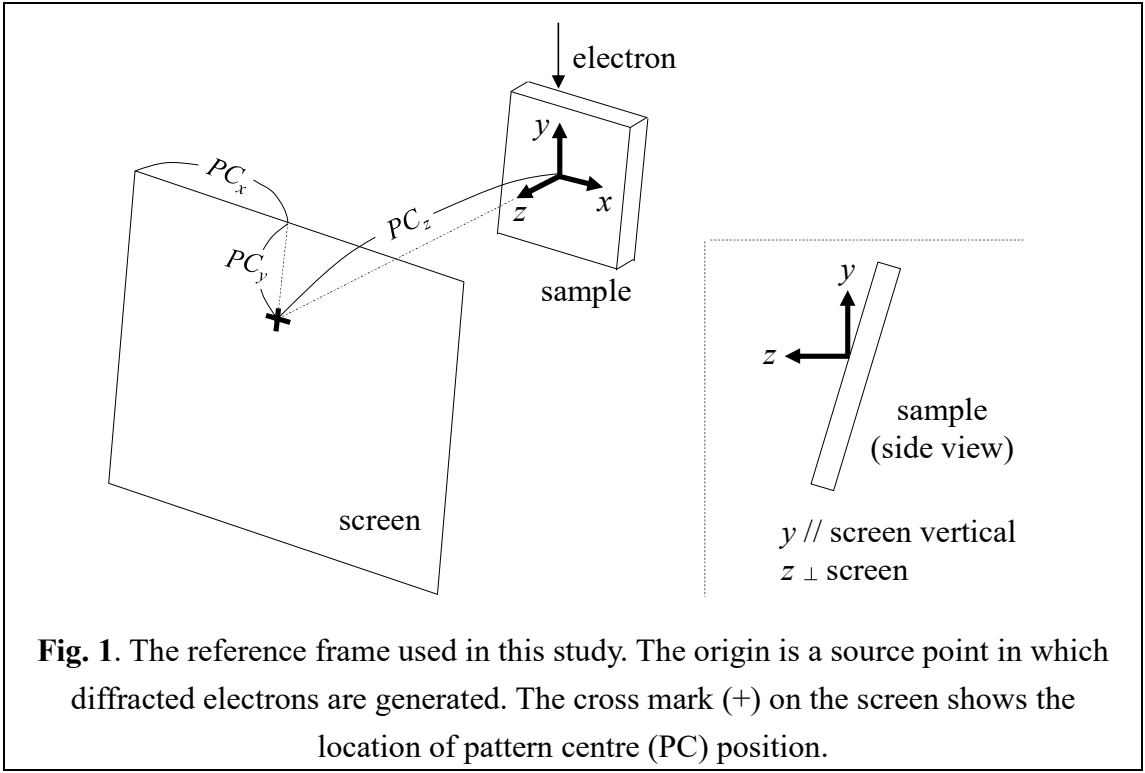

**Fig. 1**. The reference frame used in this study. The origin is a source point in which diffracted electrons are generated. The cross mark (+) on the screen shows the location of pattern centre (PC) position.

These six parameters were optimised using the DE algorithm [30, 31], which evolves a population of *Np*, *D*-dimensional target vectors (*i.e.* solution candidates) with respect to generation $g$. $D$ is the number of variables to be optimised. The target vector, $x_{i,g} = (x_{j,i,g})$, $i = 1, 2, ..., Np$, $j = 1, 2, ..., D$, $g = 1, 2, ..., g_{max}$, is initialised and evolved through the following steps:

(1) Initialisation

An initial guess for the six parameters was started by specifying both upper and lower bounds for each parameter. The first generation of each parameter, $x_{i,1} = (x_{j,i,1})$, is seeded through a random sampling within the prescribed range.

(2) Mutation

Three different vectors are picked up from the initially given parameters to mutate and recombine the population using the following relation.

$$v_{i,g} = x_{r_0,g} + F(x_{r_1,g} - x_{r_2,g}), \qquad (2)$$

where $r_0, r_1, r_2 \in \{1,..., Np\}$ that satisfies ($r_0 \neq r_1 \neq r_2 \neq i$), $v_{i,g}$ is a mutant vector and $x_{r_0,g}$, $x_{r_1,g}$ and $x_{r_2,g}$ are the three different vectors from the current generation $g$ to which the population belongs. The base vector, $x_{r_0,g}$, and the other two vectors, $x_{r_1,g}$ and $x_{r_2,g}$, can be determined in a variety of ways but in this study they were randomly chosen. The mutation factor, $F \in [0,1+]$, is a positive real number, normally in the range from 0 to 1.

(3) Crossover

To create the candidates of parameters in the next generation, DE crosses the vectors from the current generation with a mutant vector. Binomial crossover was employed in this study. Namely,

$$u_{i,g} = u_{j,i,g} = \begin{cases} v_{j,i,g} & \text{if } (\text{rand}_j \leq C_r, \text{ or } j = j_{\text{rand}}) \\ x_{j,i,g} & \text{otherwise} \end{cases}, \qquad (3)$$

where $j_{\text{rand}} \in \{1,..., D\}$ is a random parameter's index and $\text{rand}_j \in [0,1]$ is the $j$th evaluation of a uniform random number generator. The crossover probability, $C_r \in [0,1]$, controls the fraction of parameter values that are copied from the mutant vector.

(4) Selection

Either the trial vector, $\boldsymbol{u}_{i,g}$ in Eq. (3) or the vector from current generation, $\boldsymbol{x}_{i,g}$, is selected as the next generation. The criterion is whichever vector gives the lower objective function value, $f$, *i.e.* in this study the correlation coefficient, $r$, in Eq. (1) multiplied by -1.

$$x_{i,g+1} = \begin{cases} \boldsymbol{u}_{i,g} & \text{if } f(\boldsymbol{u}_{i,g}) \leq f(\boldsymbol{x}_{i,g}) \\ \boldsymbol{x}_{i,g} & \text{otherwise} \end{cases}, \qquad (4)$$

Once the new generation is selected, the steps (2)-(4) are repeated until the number of generations reaches a predetermined maximum, $g_{max}$.

In this study, the parameters for DE algorithm were set as follows: $D = 6$, $Np = 50$, $F = 0.5$, $C_r = 0.9$, $g_{max} = 100$ or $200$. This analysis is described as DE/rand/1/bin (*rand*om vector choice in mutation process, *1* vector difference addition in Eq. (2), a *bin*omial crossover). The values of $Np$, $F$, $C_r$ influence the convergence and the basic combination of the values is listed in [30, 31] with some case studies.

## 2-2. Error analysis of pattern matching EBSD analysis using differential evolution

To assess the error in retrieving PC and orientation, target EBSPs with known PC positions and orientations were generated from the master pattern. The master pattern was simulated as a stereogram with a resolution of 1901 × 1901 pixels for body centred cubic (bcc) Fe with a lattice constant of 2.866 Å using the commercial software *DynamicS* (Bruker). The acceleration voltage was set to be 20 *kV*. 10 different PC positions and orientations were selected using random number generator within the following range, *i.e.* $0° \leq \varphi_1, \varphi_2 \leq 360°$, $0° \leq \Phi \leq 180°$, $0.45 \leq PC_x \leq 0.55$, $0.30 \leq PC_y \leq 0.50$, $0.50 \leq PC_z \leq 0.90$. Using these 10 sets of parameters, gnomonic projected dynamically simulated EBSPs with a resolution of 956 × 956 pixels were extracted from the stereogram using bi-cubic interpolation. These images were Poisson-noised, distorted and binned so that the effect of these factors on the error in the calibration and indexing could be assessed. Poisson noise was applied and the noise level was assessed as a peak signal-to-noise ratio (PSNR) following the description in [34] and the barrel distortion was also applied with the distortion coefficient of $5 \times 10^{-8}$. Some of the EBSPs were binned down to 2 × 2 (478 × 478 pixels), 4 × 4 (239 × 239 pixels) and 8 × 8 (119 × 119 pixels) on a computer. For binned images, the Poisson noise was applied after the distortion application and binning were conducted. All simulated images have an image depth of 256 (8 bit).

For the DE optimisation, the range of the parameters is designated as a first step and it is necessary for the range to include true values. For this purpose, all images simulated above as target EBSPs were analysed by the commercial software *DynamicS* (Bruker), which allowed us to determine crystal orientation and PC by the undisclosed optimisation algorithm used in the software. Although the algorithm is not known for users, this gave a quick rough estimates for the parameters. One can also use the other vendor's software, such as *OIM Data Collection* supplied by EDAX for the rough estimates or the open source MATLAB code AstroEBSD [35]. Then the lower and upper bound for the DE optimisation were set as follows: the estimates by *DynamicS* ± 10° for the Euler angles ($\varphi_1$, $\Phi$, $\varphi_2$) and ± 0.05 for the PC position ($PC_x$, $PC_y$, $PC_z$).

The error in PC position and orientation was defined as a difference between the true PC position and the retrieved one, and the angle of disorientation between the true orientation and the retrieved one, respectively.

The error in orientation measurement by the classical 2 dimensional Hough transform (2DHT) based analysis was also carried out for undistorted pristine EBSPs using *OIM Data Collection* software (EDAX), assuming that the true PC position is known. The pixel size of EBSPs was reduced down to 96×96 pixels and the 2DHT was performed with a Hough resolution of 0.25°. Convolution mask size of 9×9 was used. The detected 10 Kikuchi bands were used for orientation calculation.

## 2-3. Materials and EBSD measurement

The material used in this study for EBSD measurement was interstitial free (IF) steel. The chemical composition is 0.0049C-0.004Si-0.28Al-0.003N (in wt.%). The IF steel was vacuum-melted, hot-rolled and cold-rolled down to the thickness of 1.2 mm. Following annealing at 800˚C for 10 minutes, the material was cut and polished with colloidal silica (MasterMet, Buehler). After that, in order to remove residual stress introduced by the polishing, the sample was annealed in vacuum at 700˚C for 1 hour.

The EBSPs were collected using an SEM-EBSD system (SEM: JEOL 7001F/7100F, EBSD: EDAX DigiView camera). The sample was inclined 70° from horizontal and the phosphor screen was tilted 3° from the vertical. The acceleration voltage of the field emission gun was 20 *kV* with beam current of 11~14 *nA*. The EBSPs were recorded as 8~24 bit depth tiff images with 956 × 956 pixels resolution using the circular phosphor screen. For one pattern acquisition during mapping, the integration time was about 0.5~1 s/pattern. Static background and/or dynamic background were subtracted from a raw EBSP.

Some EBSPs were recorded as a series of small rigid body rotations were applied to the sample using the SEM stage controls [1]. The compucentric stage was rotated about the sample surface normal by 0.5° with an interval of 0.1°. At each rotation angle, five patterns were recorded. The location of analysis points for the five recorded patterns at each rotation angle was within a few μm and the location of all analysis points is within the same grain during the rotation.

In order to investigate the effect of crystal orientation on the spurious strain obtained when comparing experimental patterns to dynamically simulated images, 16 EBSPs were also obtained from 16 different grains. One EBSP per grain was recorded from the 16 grains.

EBSD maps were also obtained from a strain-free region and around a pyramidal indent made with a Vickers micro-hardness tester at a load of 5 g in the IF steel. A step size of 0.2 $\mu$m over a square grid was used. Some basic features, such as inverse pole figure (IPF) map, were visualised using *TSL OIM Analysis* software (EDAX). Strain analysis was performed using in-house MATLAB code and commercially available software *CrossCourt4* (BLG Vantage). 40 or more regions of interests (ROIs) with 256 × 256 pixels size were selected over each EBSP image and cross correlation calculation was carried out in the Fourier domain after high/low pass filtering [1]. For strain analysis of the IF steel near the indent, the remapping technique [36] was applied to remove the undesired influence of lattice rotation on the elastic strain measurement. The elastic stiffness constants of $c_{11}$ = 231.5 GPa, $c_{12}$ = 135.0 GPa and $c_{44}$ = 116.0 GPa [37] for bcc Fe were used for the determination of all nine components in deformation gradient tensor assuming that the residual stress along the surface normal $\sigma_{33}$= 0.

## 3. Results & Discussion

### 3-1. Error in PC and orientation determination for simulated target EBSPs

Fig. 2(a) shows the convergence of the objective function value, $-r$ (equation 1), as a function of generation during the pattern matching through the DE algorithm. Only the minimum value at each generation is depicted. The corresponding values of the six search parameters are also displayed in Figs. 2(b)-(g). The target dynamically simulated EBSP was not noised, distorted and binned and the dotted line shows the known true values for each of the search parameters. It took 4~5 seconds for 1 generation calculation with the computer used in this study. As can be seen in Fig. 2 and Table 1, the objective function value is minimised as the generation progresses resulting in a convergence of each parameter on the true value. It is noted that between $g = 80$ and $g = 200$, the PC was changed by more than $10^{-3}$ while the objective function value $-r$ was different by only 0.00050. This is because specific combinations of the PC shifts and rotation shifts multiplied by camera length are proved to maintain the position of the Kikuchi band centres invariant [18]. Therefore, in order to determine PC and Euler angles simultaneously, tiny difference in the features of the EBSPs except for centrelines needs to be analysed, which was found to be possible by comparing entire images. It took nearly 100 generations for the PC error to become less than $10^{-3}$. Smaller number of generation would be fine to achieve good convergence if the search space (lower-upper bound range) can be narrowed, although this might lead to missing out the true value from the search space. The same optimisation was repeated 5 times to check the repeatability, which was defined as ±1 standard deviation of measured values. Despite the random nature of initial parameter generation, the repeatability was as small as ~$2.8 \times 10^{-4}$ (°) for the Euler angle and ~$1.7 \times 10^{-6}$ for the PC position. In this way, the error in PC and orientation determination was analysed for the 10 target EBSPs.

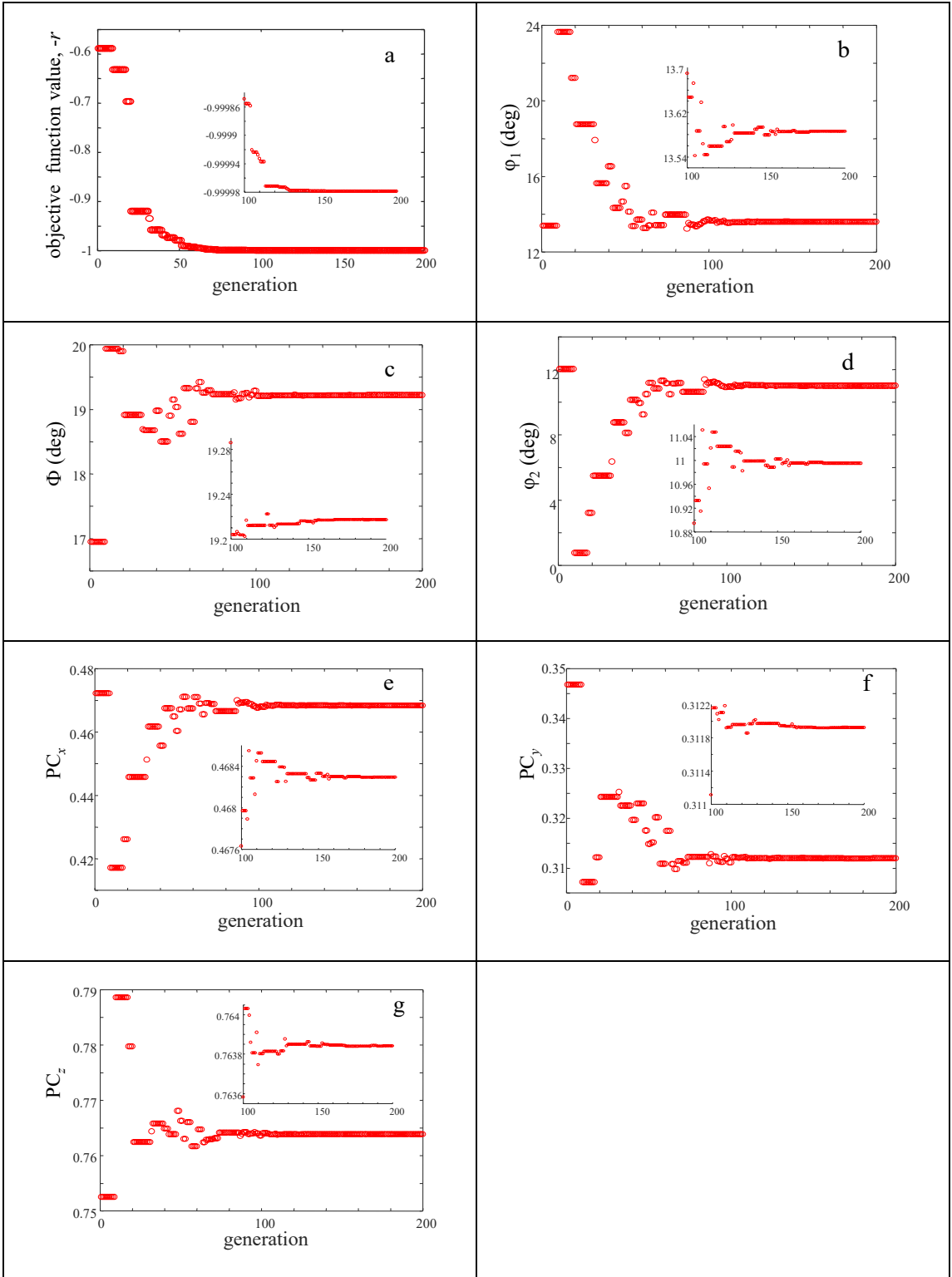

**Fig. 2**. (a) Minimisation of objective function value (negative of the cross-correlation coefficient) by the DE algorithm. (b)-(g) Optimisation of the Euler angles, (b) $\varphi_1$, (c) $\Phi$, (d) $\varphi_2$, and the PC position, (e) $PC_x$, (f) $PC_y$, (g) $PC_z$.

**Table 1**. Minimum of objective function value, $min\{-r\}$, at each generation as a function of the number of generations. The PC error $|\Delta PC_I|$ ($I = x, y, z$) and the Euler angle errors, $|\Delta\varphi_1|$, $|\Delta\Phi|$, $|\Delta\varphi_2|$ show the difference between true value and optimised value.

|  | generation, $g$ | | | | | | | | | | |
|---|---|---|---|---|---|---|---|---|---|---|---|
|  | 1 | 20 | 40 | 60 | 80 | 100 | 120 | 140 | 160 | 180 | 200 |
| $min\{-r\}$ | -0.58956 | -0.69718 | -0.96887 | -0.99323 | -0.99948 | -0.99985 | -0.99997 | -0.99998 | -0.99998 | -0.99998 | -0.99998 |
| $|\Delta PC_x|$ | $3.84 \times 10^{-3}$ | $4.22 \times 10^{-2}$ | $1.27 \times 10^{-2}$ | $9.18 \times 10^{-4}$ | $1.82 \times 10^{-3}$ | $6.61 \times 10^{-4}$ | $1.50 \times 10^{-4}$ | $3.32 \times 10^{-5}$ | $5.29 \times 10^{-6}$ | $5.00 \times 10^{-8}$ | $6.70 \times 10^{-}$ |
| $|\Delta PC_y|$ | $3.48 \times 10^{-2}$ | $2.10 \times 10^{-4}$ | $7.70 \times 10^{-3}$ | $1.08 \times 10^{-3}$ | $2.82 \times 10^{-4}$ | $8.11 \times 10^{-4}$ | $3.36 \times 10^{-5}$ | $4.47 \times 10^{-5}$ | $3.35 \times 10^{-6}$ | $4.20 \times 10^{-6}$ | $6.00 \times 10^{-}$ |
| $|\Delta PC_z|$ | $1.13 \times 10^{-2}$ | $1.59 \times 10^{-2}$ | $1.01 \times 10^{-3}$ | $2.18 \times 10^{-3}$ | $2.91 \times 10^{-4}$ | $2.59 \times 10^{-4}$ | $2.84 \times 10^{-5}$ | $7.52 \times 10^{-6}$ | $3.34 \times 10^{-6}$ | $1.39 \times 10^{-6}$ | $4.10 \times 10^{-}$ |
| $|\Delta\varphi_1|$ (deg) | 0.2097 | 7.6082 | 2.9408 | 0.1108 | 0.3682 | 0.1038 | 0.0270 | 0.0035 | $7.45 \times 10^{-4}$ | $1.57 \times 10^{-4}$ | $1.52 \times 10^{-}$ |
| $|\Delta\Phi|$ (deg) | 2.2713 | 0.6801 | 0.2430 | 0.1074 | 0.0119 | 0.0686 | 0.0051 | 0.0038 | $3.27 \times 10^{-4}$ | $2.81 \times 10^{-4}$ | $0.49 \times 10^{-}$ |
| $|\Delta\varphi_2|$ (deg) | 1.0274 | 7.7941 | 2.8914 | 0.1856 | 0.3826 | 0.1003 | 0.0287 | 0.0040 | $6.52 \times 10^{-4}$ | $2.12 \times 10^{-4}$ | $1.40 \times 10^{-}$ |

Fig. 3 shows the effect of image binning, Poisson-noise and barrel distortion on one EBSD image (PSNR ~ 20 dB). The error in PC determination with the DE algorithm was shown in Fig. 4. The error bar shows the ±1 standard deviation of measured values for 10 different cases. It is found that the PC error < $10^{-4}$ was achieved irrespective of the binning size when the pristine images were used. The standard deviation of the PC determination error was ~3 ×$10^{-5}$ at most for pristine images. When the Poisson noise with PSNR ~ 18-21 dB was applied, the PC error became bigger and reach ~$10^{-4}$ when the pattern was binned down to 2 × 2 but it is still below $10^{-3}$ for all binned images. Therefore, it can be said that the calibration using our pattern matching method is very robust against noise on the EBSPs, implying robustness against pattern collection conditions *i.e.* higher gain and lower exposure time during an actual EBSD measurement. When the barrel distortion and the Poisson noise were applied, the error in the $PC_x$ and $PC_y$ calibration was below $10^{-3}$ while the $PC_z$ error was on the order of $10^{-3}$. This is because the application of barrel distortion makes an image shrink towards the distortion centre and these image shifts are very similar to the ones caused by shortening camera length. For the precise and accurate calibration, therefore, the assessment and correction of camera lens distortion is necessary [20, 25]. It is noted that even barrel distortion coefficient can be retrieved with this DE algorithm. However, in reality, actual EBSPs can be distorted for several reasons, such as tangential distortion and image sensor distortion [25]. A comprehensive consideration of all type of distortion was not pursued in this study.

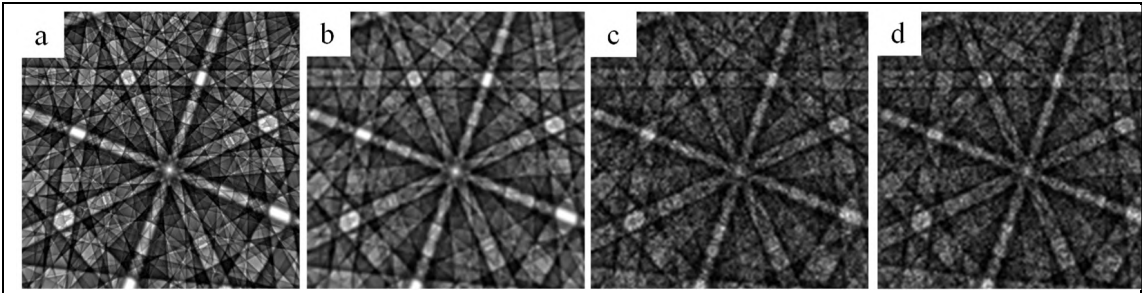

**Fig. 3**. (a) Pristine EBSP (1×1 binning). (b) 8×8 binned EBSP. (c) 8×8 binned and Poisson noise applied EBSP (PSNR ≈ 20 dB). (d) Barrel distortion, 8×8 binning and Poisson noise applied EBSP (PSNR ≈ 21 dB).

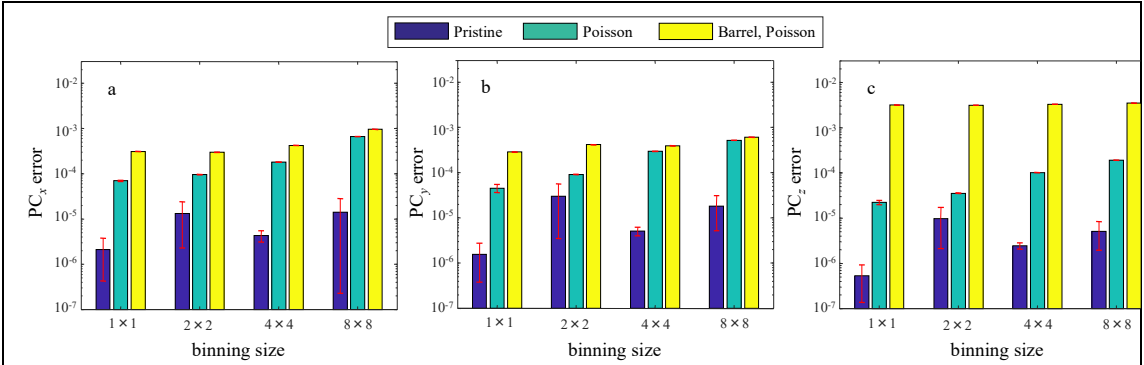

**Fig. 4**. Dependence on binning size, Poisson noise and barrel distortion of errors in the retrieval of the PC positions. (a) $PC_x$ error, (b) $PC_y$ error, (c) $PC_z$ error.

The error in the orientation determination is shown in Fig. 5. Again, the efficacy of the DE algorithm was demonstrated for pristine images with the orientation error of <0.003°. For the Poisson noise applied images, the orientation error went up to ~0.07° for 8× 8 binned EBSPs. The presence of barrel distortion increased the error up to 0.04° for 1 × 1 and 2 × 2 images.

Table 2 lists the orientation errors obtained with the conventional 2DHT analysis. The true PC position was assumed to be known. The mean orientation error is nearly 0.5°, which is close to the value reported in [26]. Therefore, it can be said that the pattern matching using the DE algorithm showed better orientation error than the 2DHT based conventional analysis algorithm [26]. It is noted that recent commercially available software supplied by Oxford Instruments showed better orientation *precision* of ~ 0.05° with the refined accuracy strategy [38] than that with conventional softwares and might show better orientation *error*. Although it would be possible to improve the orientation error obtained with 2DHT by incorporating additional refinement algorithm, this was not pursued in this study.

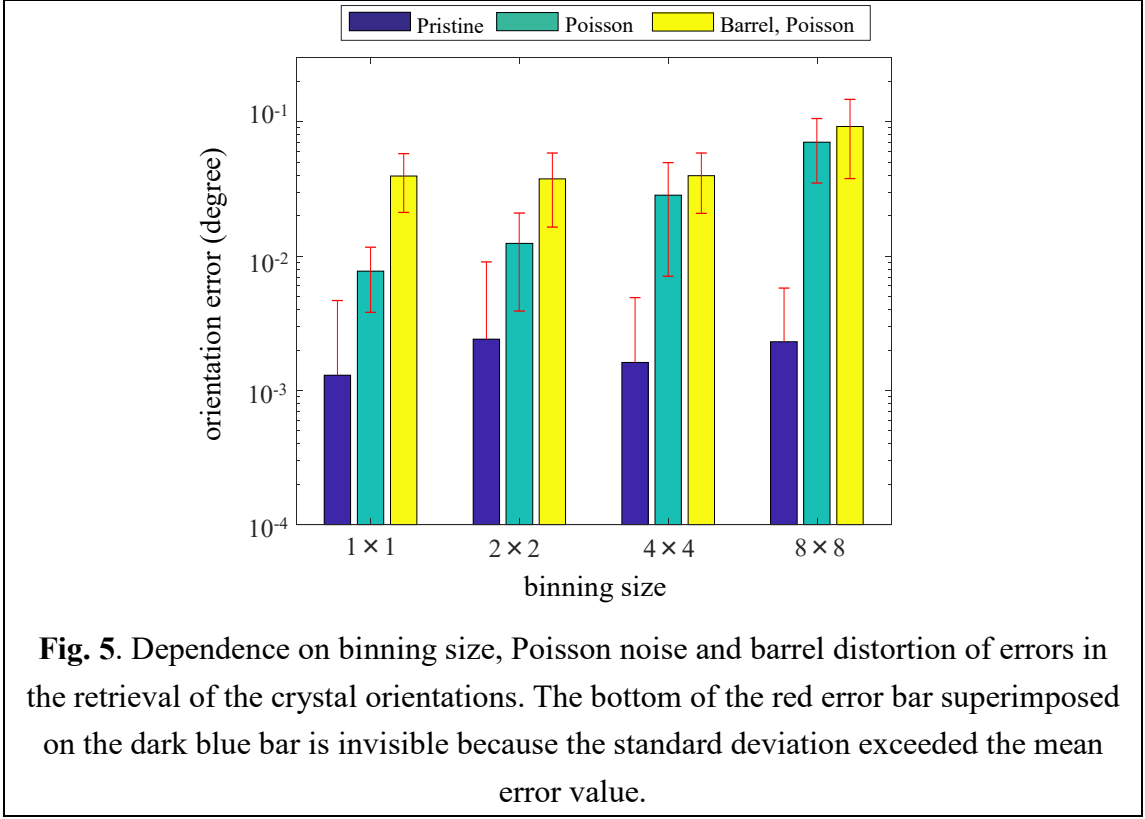

**Fig. 5**. Dependence on binning size, Poisson noise and barrel distortion of errors in the retrieval of the crystal orientations. The bottom of the red error bar superimposed on the dark blue bar is invisible because the standard deviation exceeded the mean error value.

Table 2. Retrieved Euler angle using the 2DHT based analysis.

|   | True Euler angle (deg) | | | Calculated Euler angle (deg) | | | Disori-entation (deg) |
|---|---|---|---|---|---|---|---|
|   | $\varphi_1$ | $\Phi$ | $\varphi_2$ | $\varphi_1$ | $\Phi$ | $\varphi_2$ |   |
| 1 | 13.6 | 19.2 | 11.0 | 13.5 | 19.1 | 10.9 | 0.24 |
| 2 | 259.6 | 35.6 | 117.6 | 260.0 | 35.3 | 117.3 | 1.05 |
| 3 | 17.3 | 40.3 | 25.4 | 17.4 | 40.3 | 25.1 | 0.18 |
| 4 | 323.2 | 27.8 | -4.1 | 323.5 | 27.8 | -4.3 | 0.32 |
| 5 | 148.8 | 1.0 | 180.0 | 149.7 | 1.3 | 179.0 | 0.34 |
| 6 | 106.6 | 39.8 | 277.2 | 106.1 | 39.9 | 277.7 | 0.57 |
| 7 | 215.5 | 51.3 | 124.3 | 215.5 | 51.2 | 124.4 | 0.49 |
| 8 | 274.3 | 17.3 | 50.4 | 275.0 | 17.3 | 49.7 | 0.35 |
| 9 | 300.7 | 19.6 | 47.6 | 301.8 | 19.3 | 46.8 | 0.43 |
| 10 | 123.5 | 47.9 | 222.1 | 123.8 | 47.7 | 222.0 | 0.87 |

## 3-2. Influence of background correction on PC and orientation determination for experimental EBSPs

We investigated the influence of background correction on PC and orientation determination for experimental target EBSPs. To this end, static background and/or dynamic background subtraction [39] from a raw EBSP were carried out. The static background was obtained by scanning the electron beam over large area to cover many grains at low magnifications. The dynamic background was created by blurring the original EBSP using a 2D Gaussian smoothing kernel [39] with standard deviation of 100 pixels for the full resolution raw EBSPs recorded with 956 × 956 pixels. After the static/dynamic background was subtracted, the image was rescaled so that the resulting image had a minimum intensity of zero and a maximum of 255.

Fig. 6 shows the influence of background subtraction on an experimental EBSP together with the correlation coefficient, $r$, between the target experimental and optimised dynamically simulated EBSPs. The arrows in the raw EBSP (Fig. 6(a)) show a stain or blot on the experimental pattern, which might come from pores in the phosphor particle layer. Such local defects in the detector system will not be removed using the dynamic background subtraction (Fig. 6(b)). On the other hand, static background correction removed the stain (Fig. 6(c)). The combination of static and dynamic background correction gave the best result in terms of the correlation coefficient (Fig. 6(d)). When it comes to the optimisation of PC and orientations, background correction is found to be necessary as the calculated PC/Euler angles for the raw EBSP without background correction are converged on the predetermined bound. The optimised $PC_y$ positions are slightly different between images with dynamic background correction and with static background correction by $1.1 \times 10^{-3}$ presumably because of the stain on the image. The combination of static and dynamic background correction does not alter the optimised PC position by $>1.8 \times 10^{-4}$ from the one obtained using only static background correction. Therefore, it is concluded that at least static background correction is required for quantification of elastic strain of the order of $10^{-4}$.

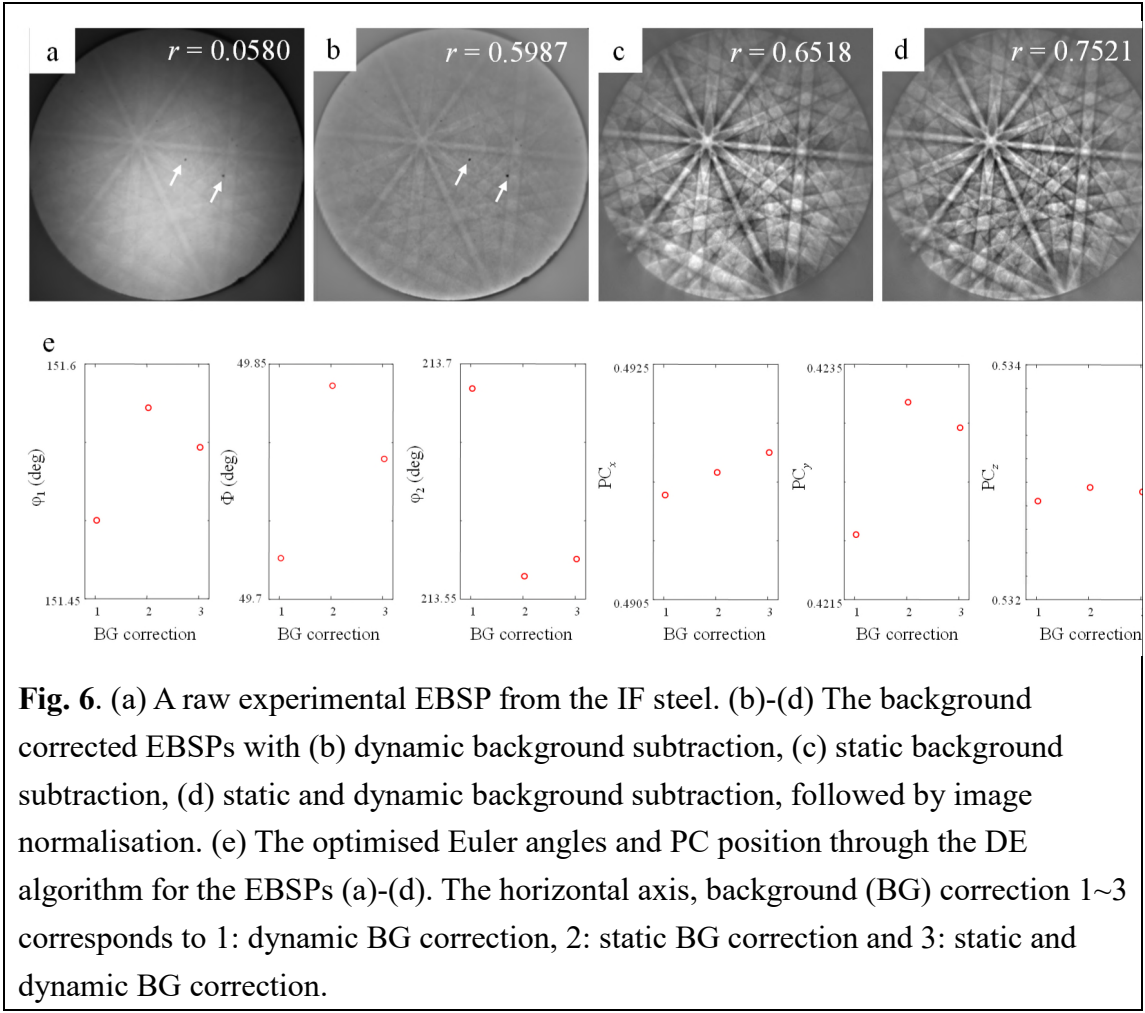

**Fig. 6**. (a) A raw experimental EBSP from the IF steel. (b)-(d) The background corrected EBSPs with (b) dynamic background subtraction, (c) static background subtraction, (d) static and dynamic background subtraction, followed by image normalisation. (e) The optimised Euler angles and PC position through the DE algorithm for the EBSPs (a)-(d). The horizontal axis, background (BG) correction 1~3 corresponds to 1: dynamic BG correction, 2: static BG correction and 3: static and dynamic BG correction.

### 3-3. Phantom strain analysis

In order to assess the accuracy of the pattern matching procedures for calibration of the detector geometry we investigate the spurious strains that can be generated when experimentally obtained EBSPs are compared with the optimised dynamically simulated EBSPs. Two experiments were conducted, namely the sample rotation experiments [1] and strain analysis around a microhardness indent, and these are described in the following sections.

#### 3-3-1. Sample rotation experiments

As stated in section 2-3, a series of EBSPs were taken after rotating the sample about the surface normal. One of the patterns from the unrotated state was selected as a reference pattern and compared to all other patterns using HR-EBSD. The calculated components of the elastic strain tensor $e_{ij}$ and rotation tensor $w_{ij}$ with respect to applied nominal rotation angle are shown in Figs. 7(a) and 7(b). Note that $e_{ij}$ and $w_{ij}$ were determined in the sample frame. The results previously obtained by Wilkinson *et al.* [1] showing a linear increase in only one of the rotation matrix components (about the sample surface normal) with respect to rotation angle was well reproduced in this figure. Fig. 7(c) shows the standard deviation of the components of $e_{ij}$ and $w_{ij}$. The better precision of strain tensor components determination of $< 10^{-4}$ compared to [1] was achieved because of the larger number of ROIs and the use of robust fitting [6].

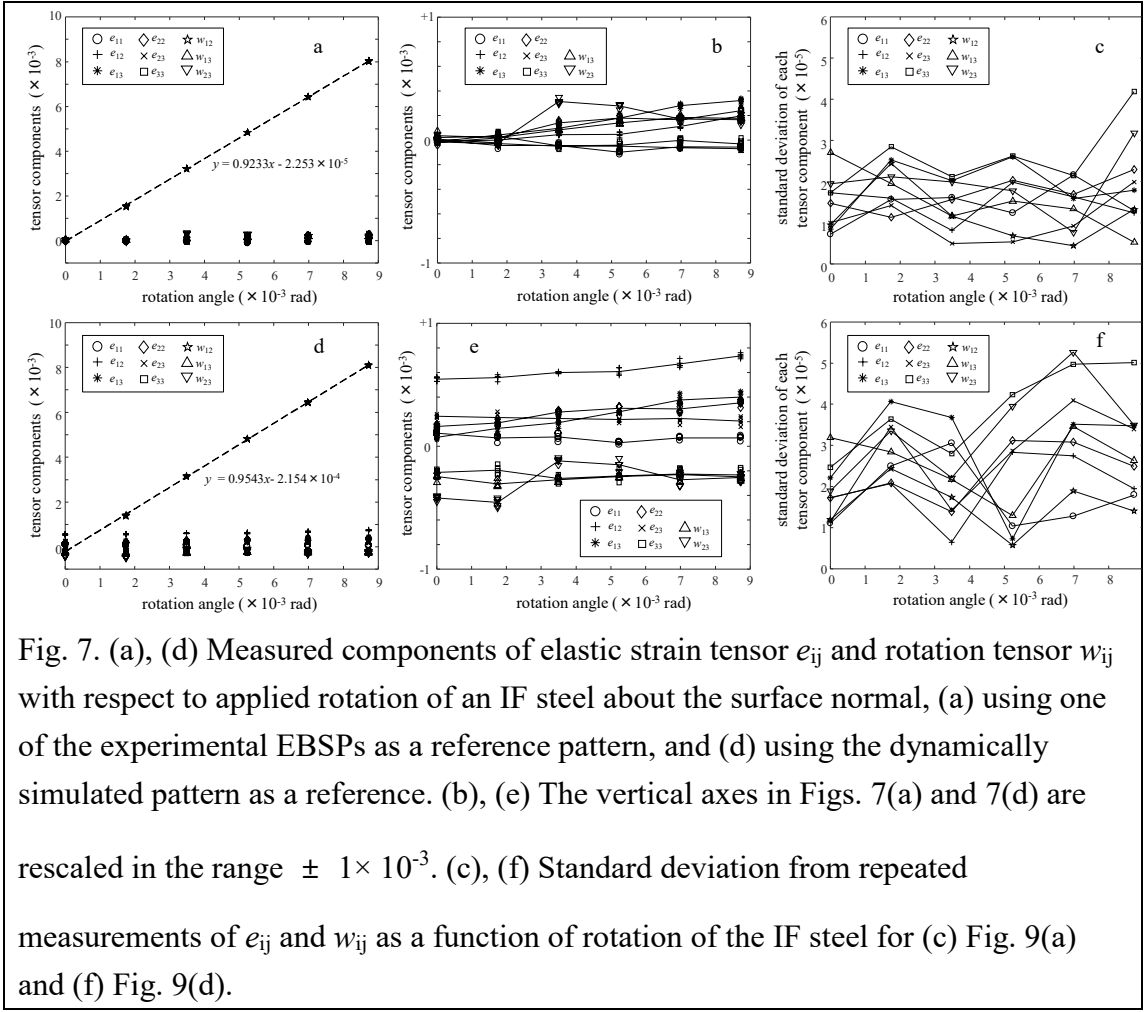

Fig. 7. (a), (d) Measured components of elastic strain tensor $e_{ij}$ and rotation tensor $w_{ij}$ with respect to applied rotation of an IF steel about the surface normal, (a) using one of the experimental EBSPs as a reference pattern, and (d) using the dynamically simulated pattern as a reference. (b), (e) The vertical axes in Figs. 7(a) and 7(d) are rescaled in the range ± 1× $10^{-3}$. (c), (f) Standard deviation from repeated measurements of $e_{ij}$ and $w_{ij}$ as a function of rotation of the IF steel for (c) Fig. 9(a) and (f) Fig. 9(d).

The experimental reference pattern was replaced with the dynamically simulated EBSP extracted from the master pattern using the optimised parameters and this used in a further cross-correlation-based analysis of the series of experimentally obtained test patterns. Figs. 7(d) and 7(e) shows the strain and rotation component results for comparison with Figs. 7(a) and 7(b). It is found that the linear increase in $w_{12}$ as a function of rotation angle and the independence of other tensor components on rotation angle are well captured. The accuracy in determining tensor components is of the order of $10^{-4}$, although the strain accuracy in $e_{12}$ stands out among all. On the other hand, the precision of tensor components determination remained < $10^{-4}$ (Fig. 7(f)), indicating that the use of dynamically simulated reference EBSPs does not deteriorate the precision of strain measurement. The standard deviation of all tensor components except for $w_{12}$ at each rotation angle was always of the order of $10^{-4}$. Therefore, it can be said that our simulation based HR-EBSD analysis still retains $10^{-4}$ order strain sensitivity.

Fig. 8 and Table 3 summarise the accuracy in determining strain/rotation tensor components obtained using dynamically simulated reference EBSPs for the other grains with 16 different crystal orientations. In Fig.8, calculated spurious strain/rotation tensor components for the other 16 grains are plotted on the IPF map. The average of the absolute values of each strain/rotation tensor component for the 16 different crystal orientations is ~6 × $10^{-4}$ with standard deviation of 5 × $10^{-4}$ at largest. Among all six independent strain tensor components, $e_{12}$ is the largest, which is consistent with Fig. 7(e). This suggests that some systematic errors are present possibly because of the distortions caused by camera lens and image sensor distortion.

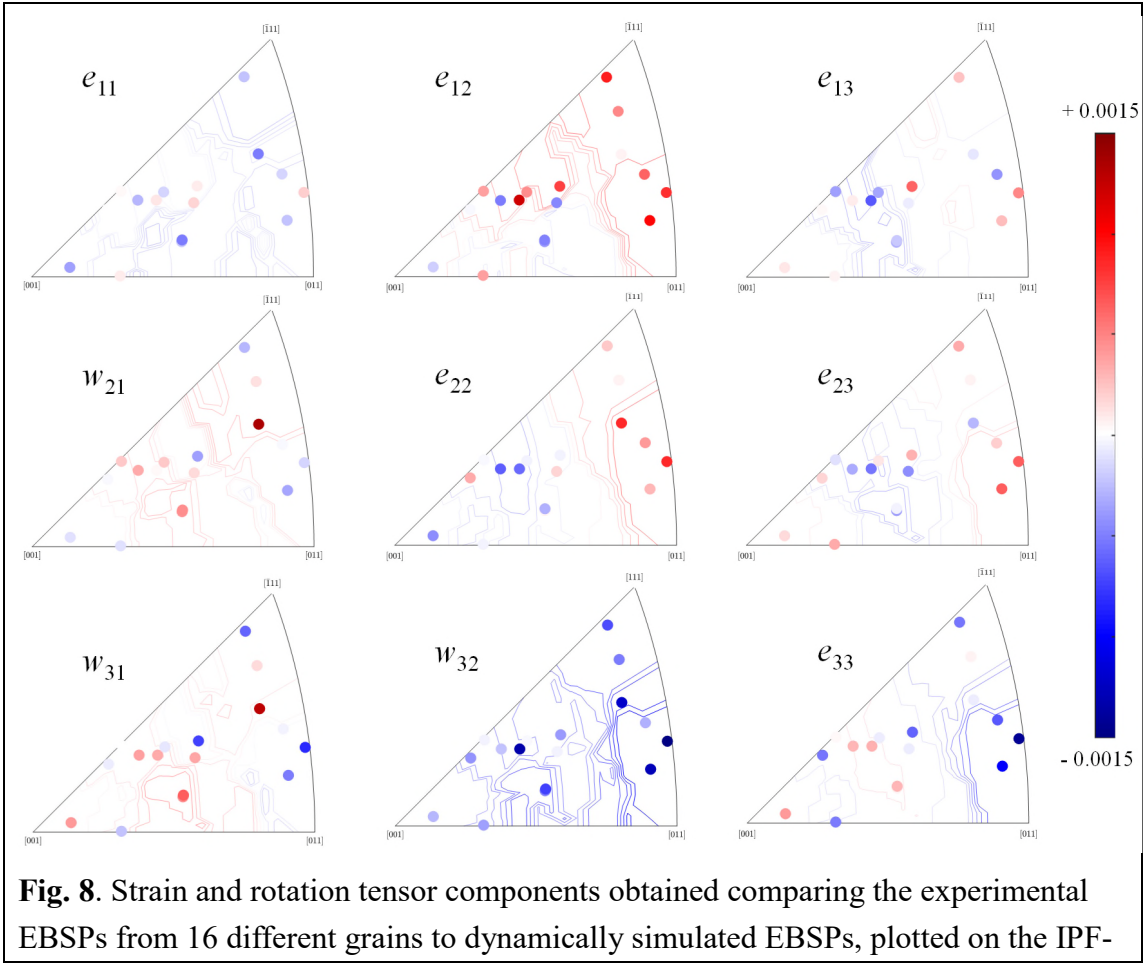

**Fig. 8**. Strain and rotation tensor components obtained comparing the experimental EBSPs from 16 different grains to dynamically simulated EBSPs, plotted on the IPF-

**Table 3**. The average of the absolute values of strain tensor $e_{ij}$ and rotation tensor $w_{ij}$, and the standard deviation of the 16 different strain tensor components shown in Fig. 8.

|  | $e_{11}$ | $e_{22}$ | $e_{33}$ | $e_{12}$ | $e_{31}$ | $e_{23}$ | $w_{21}$ | $w_{31}$ | $w_{32}$ | Frobenius norm |
|---|---|---|---|---|---|---|---|---|---|---|
| average | $1.94 \times 10^{-4}$ | $3.09 \times 10^{-4}$ | $4.01 \times 10^{-4}$ | $5.34 \times 10^{-4}$ | $2.53 \times 10^{-4}$ | $2.93 \times 10^{-4}$ | $2.78 \times 10^{-4}$ | $4.18 \times 10^{-4}$ | $5.81 \times 10^{-4}$ | $13.1 \times 10^{-4}$ |
| ±1 std | $1.50 \times 10^{-4}$ | $2.80 \times 10^{-4}$ | $3.75 \times 10^{-4}$ | $3.24 \times 10^{-4}$ | $2.04 \times 10^{-4}$ | $1.87 \times 10^{-4}$ | $3.09 \times 10^{-4}$ | $3.32 \times 10^{-4}$ | $4.91 \times 10^{-4}$ | $12.4 \times 10^{-4}$ |

This level of accuracy and precision is also confirmed by strain map analysis of the IF steel. A 5 μm × 10 μm HR-EBSD strain/rotation and mean angular error (MAE) [6] map with 5,000 analysis points was obtained using an experimental reference pattern (EBSP in Fig. 9(a)) and then the pattern was replaced with the optimised dynamically simulated reference pattern (EBSP in Fig. 9(b)). After that HR-EBSD analysis was performed again. The strain/rotation tensor components map shown in Fig. 9(b) were obtained using the simulated reference pattern. The values of all components are within ±$10^{-3}$. Table 4 lists the average of the absolute values of differences in strain tensor components |$\Delta e_{ij}$| and in rotation tensor components |$\Delta w_{ij}$| between the strain/rotation tensor obtained using experimental and simulated EBSPs. As is shown by the sample rotation experiment, the accuracy is on the order of $10^{-4}$ ($6.3 \times 10^{-4}$ at largest) with precision of ~$1 \times 10^{-4}$. This good accuracy and precision are obtained using undistorted target experimental patterns from the well annealed IF steel. In order to characterise the effect of distortion of experimental EBSPs on the accuracy and precision, the following analysis was carried out.

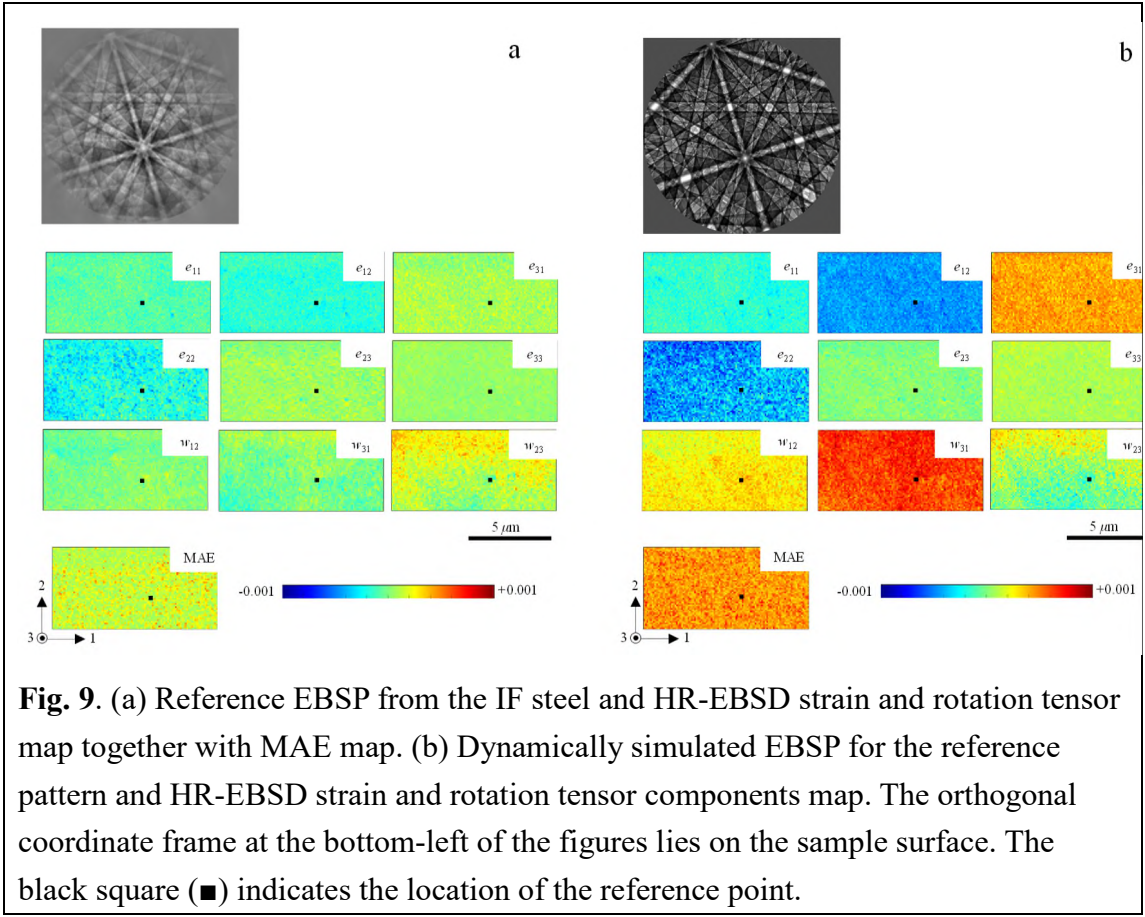

**Fig. 9**. (a) Reference EBSP from the IF steel and HR-EBSD strain and rotation tensor map together with MAE map. (b) Dynamically simulated EBSP for the reference pattern and HR-EBSD strain and rotation tensor components map. The orthogonal coordinate frame at the bottom-left of the figures lies on the sample surface. The black square (■) indicates the location of the reference point.

**Table 4**. The absolute value of the strain and rotation tensor difference, $|\Delta e_{ij}|$ and $|\Delta w_{ij}|$, between HR-EBSD results using experimental reference pattern and dynamically simulated pattern when analysing strain free region in annealed IF steel.

|  | $|\Delta e_{11}|$ | $|\Delta e_{22}|$ | $|\Delta e_{33}|$ | $|\Delta e_{12}|$ | $|\Delta e_{31}|$ | $|\Delta e_{23}|$ | $|\Delta w_{21}|$ | $|\Delta w_{31}|$ | $|\Delta w_{32}|$ |
|---|---|---|---|---|---|---|---|---|---|
| average | $0.83 \times 10^{-4}$ | $2.20 \times 10^{-4}$ | $0.47 \times 10^{-4}$ | $2.80 \times 10^{-4}$ | $3.45 \times 10^{-4}$ | $0.50 \times 10^{-4}$ | $2.70 \times 10^{-4}$ | $6.30 \times 10^{-4}$ | $1.06 \times 10^{-4}$ |
| ±1 std | $0.33 \times 10^{-4}$ | $1.10 \times 10^{-4}$ | $0.29 \times 10^{-4}$ | $0.44 \times 10^{-4}$ | $0.54 \times 10^{-4}$ | $0.31 \times 10^{-4}$ | $0.57 \times 10^{-4}$ | $0.64 \times 10^{-4}$ | $0.67 \times 10^{-4}$ |

### 3-3-2. Strain analysis near the indent in IF steel

The strain near the indent in the IF steel was investigated with the use of experimental and simulated reference patterns. Fig. 10(a) shows an SEM image of the indent in the IF steel. The surface was nearly parallel to (111) plane (Fig. 10(b)). The SEM image was also obtained after the sample stage was inclined 70° to horizontal as shown in Fig. 10(c), and in Fig. 10(d) after a 180° rotation of the sample to show the other side of the indent.    The edges of the indent clearly curve towards the interior of the indent impression indicating that the plastic flow around the indent is characteristic of 'sink-in' rather than 'pile-up' [40].

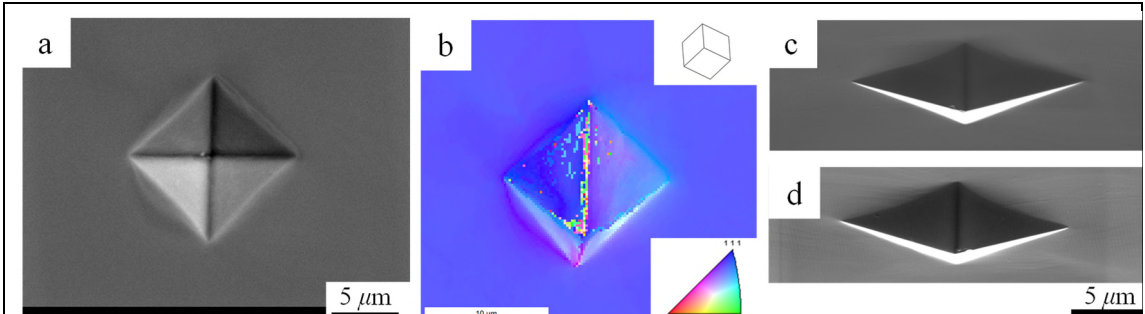

Fig. 10. (a) SEM image of the indent in the IF steel. (b) IPF-ND (surface Normal Direction) map around the indent, (c) SEM image taken after 70° sample stage tilting to horizontal, showing the curvature of sides of square edge. (d) SEM image after 180° rotation from Fig. 10(c).

Lattice strain and rotation fields were obtained from the HR-EBSD analysis using the experimental reference pattern obtained from Point A (Fig. 11(a)) located near the top left corner of the map, away from the indent and where the crystal can be assumed to be strain-free, or at least containing only very small strains. The stress fields in the same area are available in the Supplementary section. The lattice rotations are seen to be markedly larger than the residual lattice strains (note different colour-scales are used).

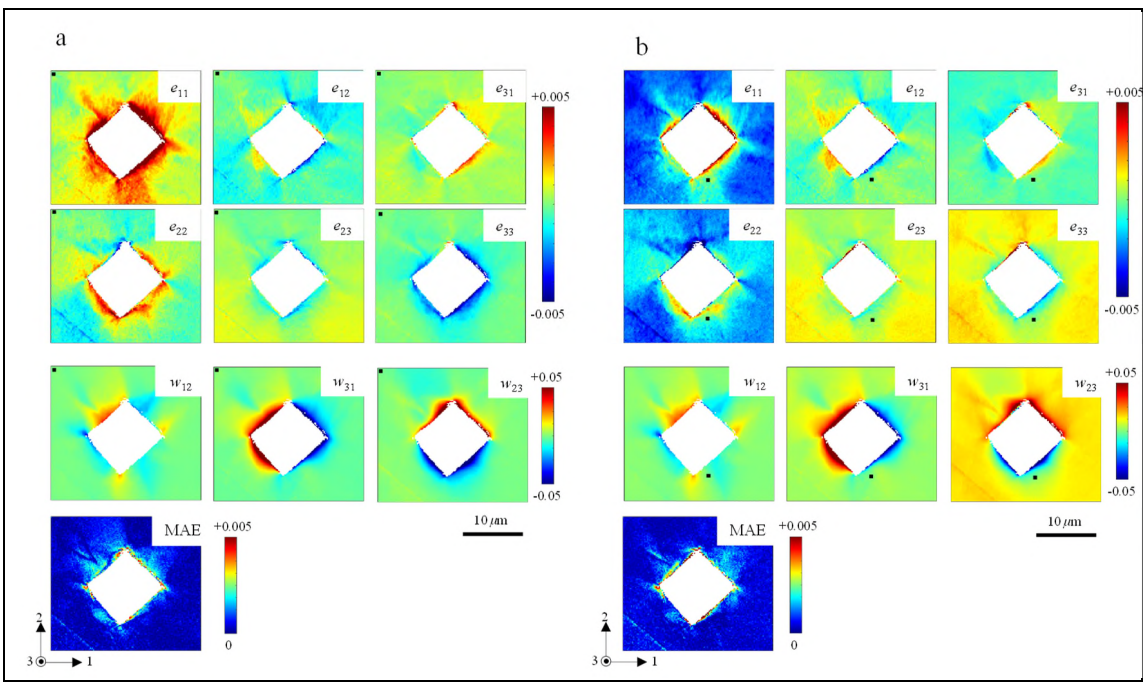

**Fig. 11**. Strain and rotation tensor $e_{ij}$, $w_{ij}$ map around the indent using the experimental reference EBSP either (a) from Point A (top left corner) or (b) from Point B (lower right of the indent). The orthogonal coordinate frame at the bottom-left of the figures lies on the sample surface. The location of the reference point is marked as the black square (■). The pixel with MAE > 5 × 10$^{-3}$ is shown in white colour.

A second HR-EBSD analysis was undertaken on the same set of EBSPs, with the reference point was changed from Point A to Point B which is sufficiently close to the indent edge, as shown by the black square in Fig. 11(b), to have significant lattice strains. It is very evident from comparison of Fig. 11(a) and Fig. 11(b) that HR-EBSD measures relative strain, not absolute strain. This is most clearly seen in the normal strains which vary considerably between the two reference points and generate a significant offset in the apparent relative strain maps. Therefore, when experimental reference pattern is used, the analysed strain is shown to be dependent on the location of reference point [41] and the strain state at that point. This is a critical issue with conventional HR-EBSD, as in most applications the strain fields are too complex to identify a position where the strain state is known *a priori*.

Fig. 12 shows experimental reference patterns from Points A (Fig. 12(a)) and B (Fig. 12(c)) used for the above HR-EBSD analysis. The pattern from Point A was replaced with the dynamically simulated EBSP (Fig. 12(b)) obtained after pattern matching to obtain the best fit crystal orientation and PC position using the DE optimisation procedures. The pattern from Point B was also replaced with a second dynamically simulated EBSP (Fig. 12(d)), although in this case only the Euler angles were optimised while the PC position was obtained from the PC position at Point A corrected for the beam shift from Point A to B. This is because deterioration in the PC determination error is expected to occur through the pattern matching between undistorted dynamically simulated EBSP and distorted experimental one.

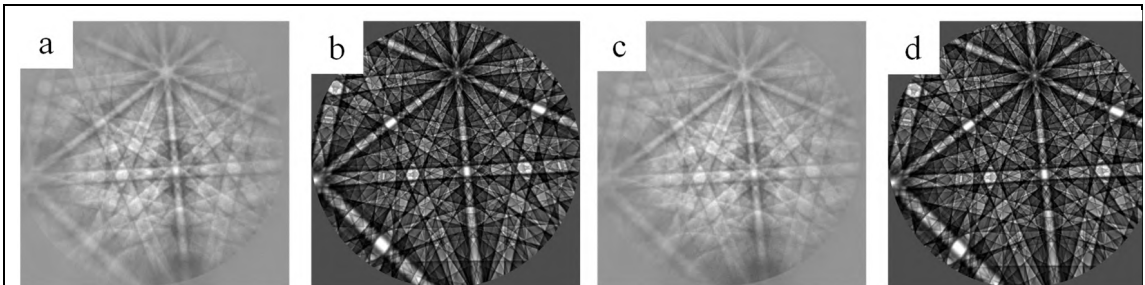

**Fig. 12**. Experimental reference EBSPs from (a) Point A and (c) Point B. These patterns were replaced with dynamically simulated EBSPs (b) for Point A and (d) for Point B.

The strain distribution obtained from HR-EBSD analysis using each of these two dynamically simulated reference patterns is shown in Fig. 13. It is obvious that the resulting two lattice strain field distributions in Fig. 13 obtained using the strain free simulated reference patterns are much more similar than those in Fig. 11(b) obtained using the experimental patterns which contain distortions from (unknown) lattice strain. The tensile residual strain in $e_{11}$ and $e_{22}$ near the indent is obtained using not only the reference pattern from Point A but also the pattern from Point B. The strain difference between Figs. 13(a) and 13(b) is shown in detail in Fig. S2 in Supplementary section. Therefore, the efficacy of the use of dynamically simulated EBSP for absolute strain analysis is verified. To check the strain determination accuracy, strain values at each strain map pixel in Figs 11(a) and 13 were compared to each other.

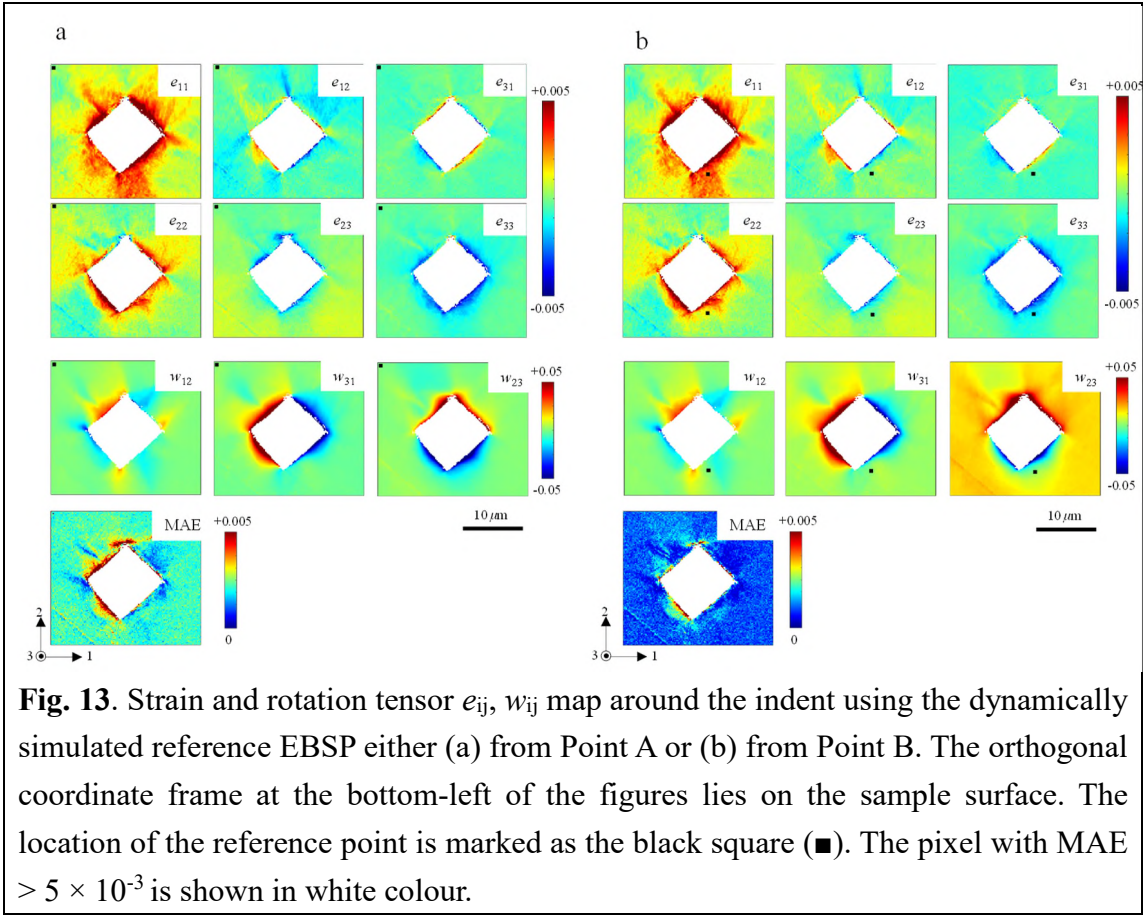

**Fig. 13**. Strain and rotation tensor $e_{ij}$, $w_{ij}$ map around the indent using the dynamically simulated reference EBSP either (a) from Point A or (b) from Point B. The orthogonal coordinate frame at the bottom-left of the figures lies on the sample surface. The location of the reference point is marked as the black square (■). The pixel with MAE $> 5 \times 10^{-3}$ is shown in white colour.

Table 5 shows that the average of the absolute values of strain/rotation component differences and its precision for each component are less than $< 10^{-3}$. The accuracy of $e_{31}$ stands out as it is significantly higher than the accuracy of the other components.

Compared to the results in Table 4, deterioration of accuracy and precision is observed. This is probably because of the pattern blurring caused by the extensive plastic deformation. The $\Delta e_{ij}$ (Fig. 11(a) vs Fig.13(a)) map (Fig. 14) displays the spatial variation in strain measurement accuracy, showing the variation occurs mainly near the indent edge. Thus when highly deformed materials are to be investigated by simulation based HR-EBSD, the accuracy and precision would be influenced according to the extent of local deformation. However, the fact that $< 1 \times 10^{-3}$ accuracy and precision are achieved in Fig. 13(b) using the reference pattern from distorted point indicates that a simulated strain-free reference point can be generated for any point in a map as long as relatively clear patterns are available.

**Table 5**. The average of the absolute values of strain and rotation tensor difference, $|\Delta e_{ij}|$, between HR-EBSD results due to the use of different reference patterns. 'Exp' and 'Sim' in the first column corresponds to experimental and dynamically simulated reference patterns respectively. Point A and Point B show the reference points from which the reference patterns are taken.

|  |  | $|\Delta e_{11}|$ | $|\Delta e_{22}|$ | $|\Delta e_{33}|$ | $|\Delta e_{12}|$ | $|\Delta e_{31}|$ | $|\Delta e_{23}|$ |
|---|---|---|---|---|---|---|---|
| \|Exp(Point A) - Sim(Point A)\| | average | $1.87 \times 10^{-4}$ | $6.11 \times 10^{-4}$ | $1.96 \times 10^{-4}$ | $1.40 \times 10^{-4}$ | $7.54 \times 10^{-4}$ | $1.95 \times 10^{-4}$ |
| | ±1 std | $3.69 \times 10^{-4}$ | $4.50 \times 10^{-4}$ | $1.80 \times 10^{-4}$ | $1.85 \times 10^{-4}$ | $3.96 \times 10^{-4}$ | $2.66 \times 10^{-4}$ |
| \|Exp(Point A) - Sim(Point B)\| | average | $3.08 \times 10^{-4}$ | $5.23 \times 10^{-4}$ | $1.27 \times 10^{-4}$ | $5.82 \times 10^{-4}$ | $9.64 \times 10^{-4}$ | $1.78 \times 10^{-4}$ |
| | ±1 std | $3.58 \times 10^{-4}$ | $4.82 \times 10^{-4}$ | $1.69 \times 10^{-4}$ | $2.53 \times 10^{-4}$ | $3.87 \times 10^{-4}$ | $1.80 \times 10^{-4}$ |
| \|Sim(Point A) - Sim(Point B)\| | average | $3.51 \times 10^{-4}$ | $1.54 \times 10^{-4}$ | $1.14 \times 10^{-4}$ | $5.45 \times 10^{-4}$ | $2.43 \times 10^{-4}$ | $6.41 \times 10^{-4}$ |
| | ±1 std | $2.52 \times 10^{-4}$ | $3.05 \times 10^{-4}$ | $1.40 \times 10^{-4}$ | $1.24 \times 10^{-4}$ | $2.47 \times 10^{-4}$ | $1.77 \times 10^{-4}$ |

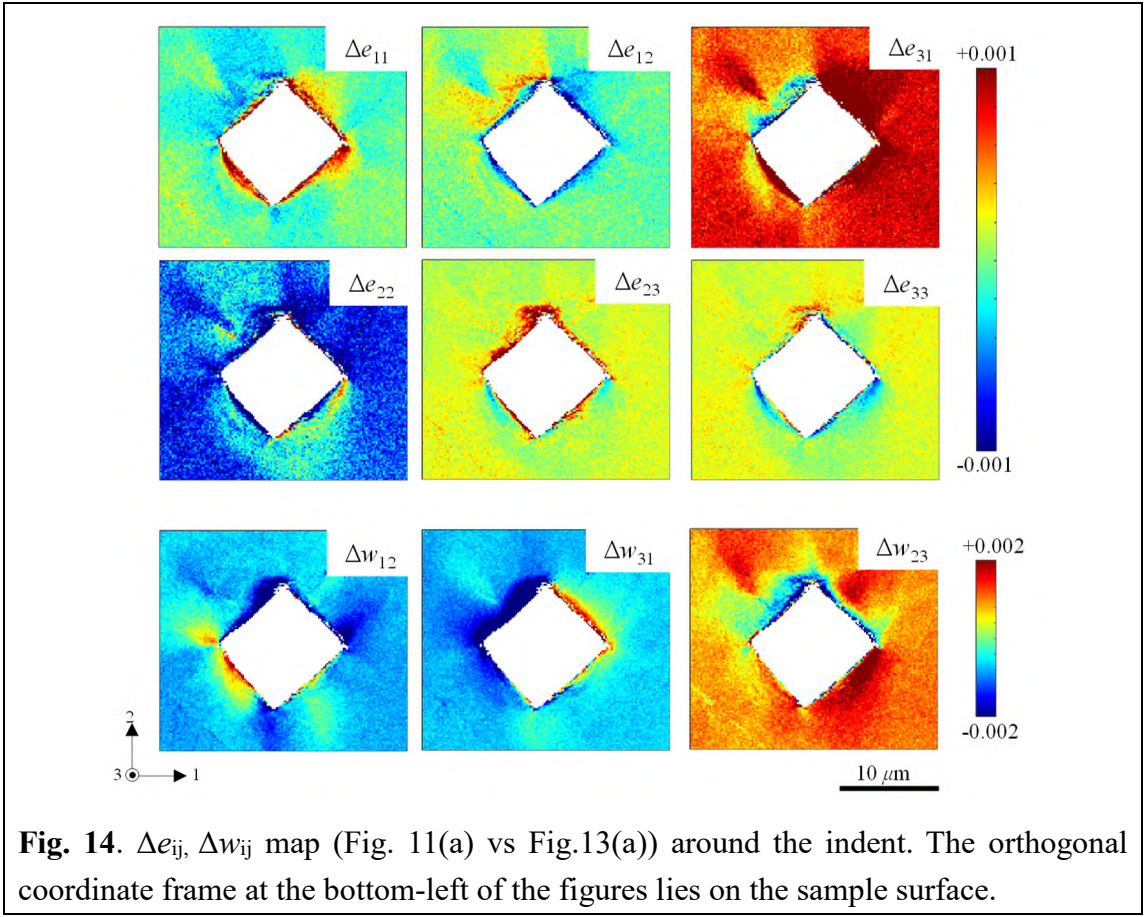

**Fig. 14**. $\Delta e_{ij}$, $\Delta w_{ij}$ map (Fig. 11(a) vs Fig.13(a)) around the indent. The orthogonal coordinate frame at the bottom-left of the figures lies on the sample surface.

The measured ~$10^{-4}$ order accuracy is surprising considering the result that the typical barrel distortion with a distortion coefficient of $1 \times 10^{-7}$ can cause strain errors of ~$3 \times 10^{-3}$ [17]. There are two possible reasons for this. Firstly, the barrel distortion coefficient for the camera used in our study might be smaller than $1 \times 10^{-7}$. Britton *et al* reported that the distortion coefficient $< 0.3 \times 10^{-7}$ is necessary to achieve $< 10^{-3}$ accuracy in strain measurement when comparing unstrained reference pattern to barrel distorted test pattern [17]. This means the strain error/accuracy obtained in this study is not always applicable to the EBSD measurement using different detector system as the distortion coefficient differs from camera to camera [25]. Secondly, the $PC_z$ (camera length) is optimised not on the true value (see Fig. 4(c)) but on the value which gives the minimum strain error. In this case, the strain error caused by the presence of lens distortion might be compensated to some extent by the strain error introduced by $PC_z$ error. In either case, in order to explore the possibility to further reduce strain error, the assessment (and correction, if necessary) of lens distortion/image sensor distortion is a key requirement. An alternative approach is the use of direct detection system [42] to avoid the lens distortion related issue.

Another issue with our pattern matching approach is the use of dynamically simulated EBSPs undistorted by any assumed non-zero strain state to be compared to target EBSPs. This means that an undistorted experimental EBSP taken from strain-free region is likely to lead to the best possible result from the calibration through the pattern matching. Incorporating deformation gradient tensor into the dynamical simulation of multiple master patterns within the DE algorithm is not a plausible solution with current computer power [16]. We note that for the simulated pattern used for the strained point B in Fig. 13(b) the pattern matching had the PC position fixed and searched only for a crystal orientation. An additional accurate calibration technique is being pursued by the present authors for use in situations when a strain-free region is not available near the analysis area.

During the review process of this paper, a novel approach for absolute strain/stress analysis within HR-EBSD analysis without using a simulated reference pattern was demonstrated in the papers [43, 44], exploiting crystal symmetry and plane-stress condition. It is claimed that the strain determination error is less than $10^{-4}$. However this was achieved using dynamically simulated patterns with known distortion. Some factors observed in actual EBSPs, such as an asymmetry in the intensity profile across Kikuchi bands, the presence of optical distortion and strain variation within EBSD analysis volume, would deteriorate the strain measurement error. Further validation using experimental EBSPs is required.

## 4. Summary and Conclusion

The calibration and indexing of a target EBSP through pattern matching with dynamically simulated EBSPs within an iterative optimisation using a differential evolution (DE) scheme was demonstrated. The pattern matching with the DE algorithm allows for the very accurate, precise and simultaneous PC/orientation determination with good robustness against noise. The presence of barrel distortion influenced the error in determining camera length ($PC_z$) as the image change caused by camera length shift and by barrel distortion is very similar. The experimental EBSPs to be compared to dynamically simulated patterns should be background corrected.

The efficacy of the use of dynamically simulated EBSPs as reference patterns within HR-EBSD strain analysis was verified through phantom strain analysis. Strain measurement accuracy of less than $1 \times 10^{-3}$ was achieved for IF steels and was at its worst when patterns in deformed regions close to an indent were used for the pattern matching. In deformation free crystals, phantom strains were typically rather smaller at ~6 × $10^{-4}$. Further improvement in the accuracy should be achieved by taking lens distortion/image sensor distortion into account. In addition to this, one of the biggest obstacles to be overcome is the PC determination through pattern matching when analysing the sample with no strain-free region near the analysis area. Both issues are being addressed for the application of simulation based HR-EBSD to a wider engineering problems.

The pattern matching approach provides a more accurate route for detector geometry calibration than existing methods. Sufficient accuracy is achieved in PC quantification for strain free reference patterns to be simulated using dynamical diffraction theory and used successfully in HR-EBSD strain mapping. Precision in the measurements still exceeds the accuracy which continues to be limited by knowledge of the camera geometry and optical aberrations.

# Supplementary Material

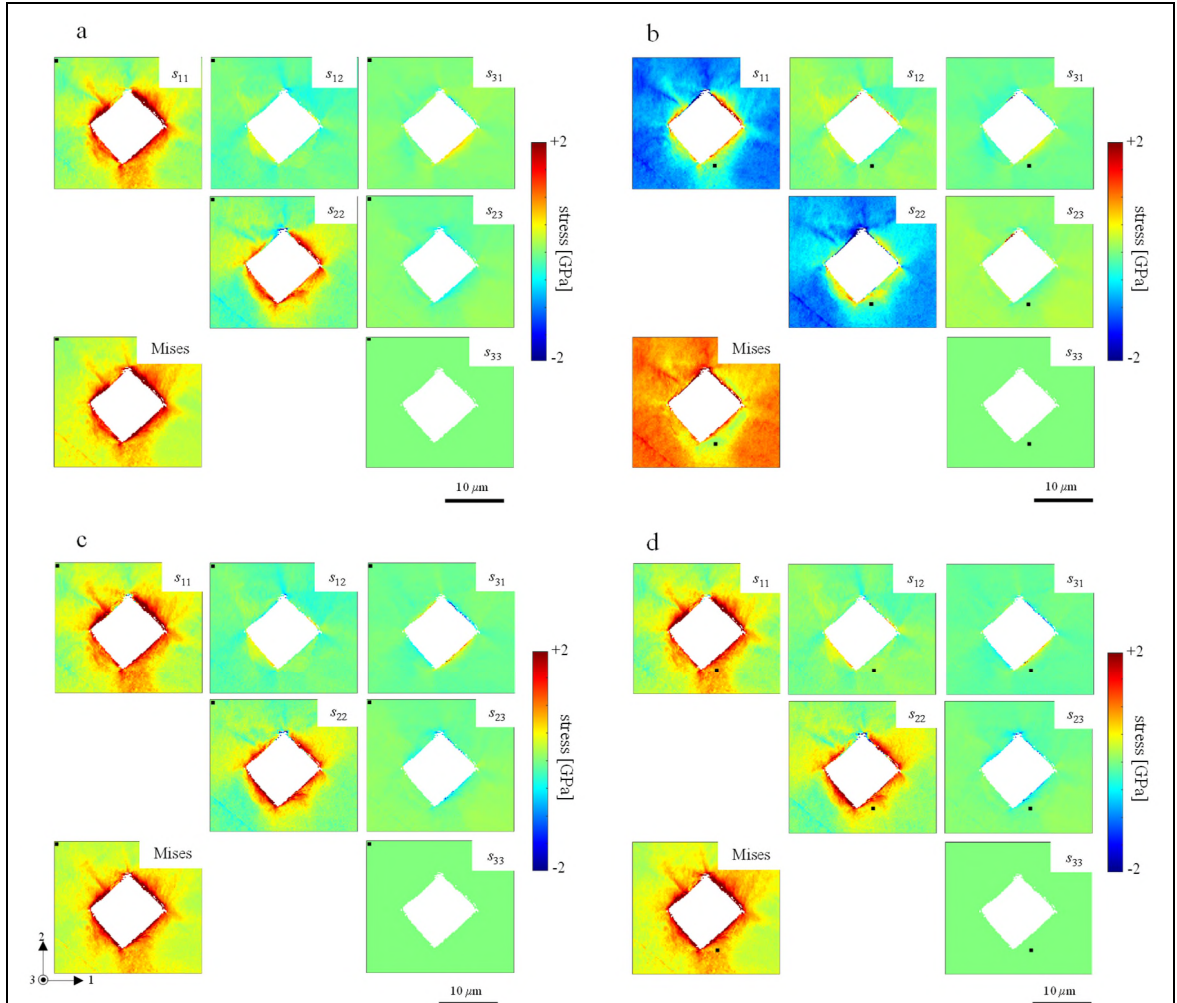

**Fig. S1.** Stress tensor $s_{ij}$ and Mises stress map around the indent using the experimental reference EBSP either (a) from Point A (top left corner) or (b) from Point B (lower right of the indent). $s_{ij}$ map in the same area using the dynamically simulated pattern either (c) from Point A or (d) from Point B. The orthogonal coordinate frame at the bottom-left of the figures lies on the sample surface. The location of the reference point is marked as the black square (■). The pixel with MAE > $5 \times 10^{-3}$ is shown in white colour.

**Table S1.** The average of the absolute values of stress tensor difference, $|\Delta s_{ij}|$, between HR-EBSD results due to the use of different reference patterns. The figures in the table are in [GPa]. 'Exp' and 'Sim' in the first column corresponds to experimental and dynamically simulated reference patterns respectively. Point A and Point B show the reference points from which the reference patterns are taken.

|  |  | $|\Delta s_{11}|$ | $|\Delta s_{22}|$ | $|\Delta s_{12}|$ | $|\Delta s_{31}|$ | $|\Delta s_{23}|$ | $|\Delta \text{Mises}|$ |
|---|---|---|---|---|---|---|---|
| |Exp(Point A) − Sim(Point A)| | average | 0.119 | 0.134 | 0.027 | 0.096 | 0.030 | 0.086 |
|  | ±1 std | 0.101 | 0.143 | 0.035 | 0.061 | 0.040 | 0.101 |
| |Exp(Point A) − Sim(Point B)| | average | 0.054 | 0.086 | 0.105 | 0.127 | 0.024 | 0.102 |
|  | ±1 std | 0.086 | 0.142 | 0.047 | 0.058 | 0.032 | 0.105 |
| |Sim(Point A) − Sim(Point B)| | average | 0.095 | 0.077 | 0.100 | 0.037 | 0.041 | 0.093 |
|  | ±1 std | 0.080 | 0.098 | 0.029 | 0.039 | 0.027 | 0.090 |

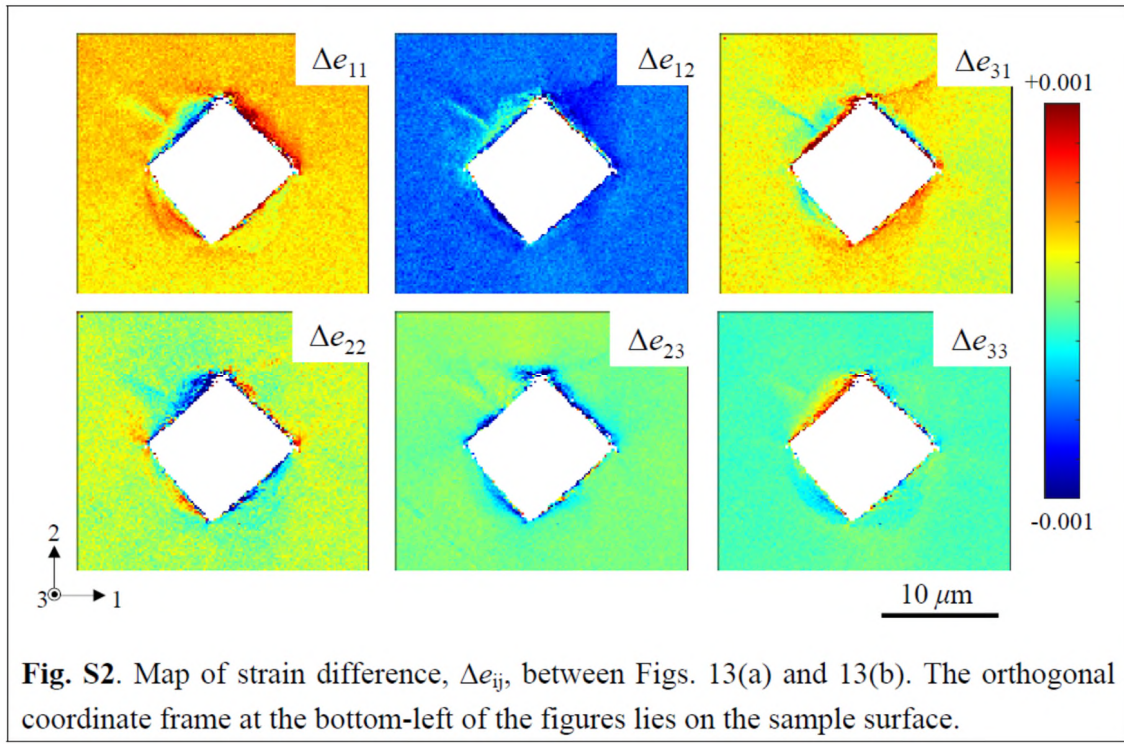

**Fig. S2.** Map of strain difference, $\Delta e_{ij}$, between Figs. 13(a) and 13(b). The orthogonal coordinate frame at the bottom-left of the figures lies on the sample surface.